# EVOLUTION OF STRUCTURE OF SOME BINARY GROUP-BASED N-BIT COMPARATOR, N-TO-$2^N$ DECODER BY REVERSIBLE TECHNIQUE

Neeraj Kumar Misra[1], Subodh Wairya[2] and Vinod Kumar Singh[3]

Department of Electronics Engineering,
Institute of Engineering and Technology, Lucknow, India

## ABSTRACT

*Reversible logic has attracted substantial interest due to its low power consumption which is the main concern of low power VLSI circuit design. In this paper, a novel 4x4 reversible gate called inventive gate has been introduced and using this gate 1-bit, 2-bit, 8-bit, 32-bit and n-bit group-based reversible comparator have been constructed with low value of reversible parameters. The MOS transistor realizations of 1-bit, 2- bit, and 8-bit of reversible comparator are also presented and finding power, delay and power delay product (PDP) with appropriate aspect ratio W/L. Novel inventive gate has the ability to use as an n-to-$2^n$ decoder. Different proposed novel reversible circuit design style is compared with the existing ones. The relative results shows that the novel reversible gate wide utility, group-based reversible comparator outperforms the present design style in terms of number of gates, garbage outputs and constant input.*

## KEYWORDS

*Reversible Logic, Inventive Gate, Garbage Output, Constant input, Full subtraction, n-bit reversible comparator, Reversible decoder etc.*

## 1. INTRODUCTION

In low power VLSI circuit, planning of power is one of important aspects. According to Landauer [1] in 1960 demonstrate that single bit of information loss generate at least KT ln2 J/K of energy where K is the Boltzmann constant ($1.38 \times 10^{-23}$ J/K) and T is the absolute temperature. Reversible circuits are totally different from irreversible circuits. In reversible logic no bits is loss the circuit that doesn't loss information is reversible.

C.H Bennet [2] in 1973 proves that KT ln2 joule of energy wouldn't be dissipated if the reversible circuit consist of reversible gates only. Thus reversible logic operations do not loss information and dissipate less heat also as power. Thus reversible logic is probably going to be in demand in high speed power aware circuits. A reversible circuit planning has following important attributes such as Garbage output, number of reversible gates, constant inputs, all should be minimum for efficient reversible circuits.

Comparator has wide applications in Analog and digital circuits, Analog to digital (A/D) converters, Level shifter, and communication system etc. It compares the 2-number of several

  9

bits. In this paper introduce group-based n-bit reversible comparator [6,9,11] and n-to-$2^n$ reversible decoder with low value of reversible parameters with the help of various lemmas.

This paper is organized with the following sections: Section 2 and 3 discuss basic definition of reversible logic; Section 4 discuss the past work; Section 5 discuss the utility and design issue of novel 4x4 inventive gate; Section 6 Planning of low value style reversible comparator subsection of 6.1 introduce novel 1-bit comparator cell, subsection 6.2 and 6.3 is novel match, larger and smaller comparator cell design; Subsection 6.4, 6.5, 6.6 and 6.7 for 2-bit, 8-bit, 32-bit and n- bit group-based reversible comparator design respectively; Section 7 Implement all comparator cell in MOS transistor with minimum MOS transistor count. Section 8 categories as simulation result of comparator. Finally, the paper is concluded and future work with Section 9.

## 2. BASIC DEFINITIONS OF REVERSIBLE LOGIC

In this section, we introduce the essential definitions of reversible logic which are relevant with this research work.

**Definition 2.1** A reversible gate is a Z - input and Z - output that generate a unique output pattern for each possible input pattern.

**Definition 2.2** In Reversible logic output and input is equal in number. And unwanted output is called garbage output it should be minimum as possible.

## 3. PARAMETER OPTIMIZED FOR DESIGNING EFFICIENT REVERSIBLE CIRCUITS

The main challenge of designing efficient reversible circuits is to optimize the different reversible parameters which result the circuit design is costly. The most necessary parameters which have dominant in efficient reversible logic circuits are:

### 3.1 Garbage output

Unutilized or unwanted output of reversible circuit is called garbage output. It should be kept minimum as possible.

### 3.2 Constant Input

Constant bits are additional inputs that are not part of the original specification. These bits are added in hopes to reduce the circuit complexity or realize a reversible function. They come in the form of a constant logic 1 or 0. It is ideal to keep in minimal.

### 3.3 Few Reversible gates utilized

In this subsection, we present few reversible gates that are used to design for planning different types of reversible circuit. First TR (3×3) utilize input (A, B, C) and carry output (P=A, Q= $A \oplus B$, $R = AB' \oplus C$, Second BME gate(4X4) utilize input (A, B, C, D) and carry output (P=A, $Q = AB \oplus C$, $R = AD \oplus C$) and $S = A'B \oplus C \oplus D$, Third Feynman gate (FG) 2x2 utilize input (A, B) and carry output (P=A, $R = A \oplus B$), Fourth Peres gate (PG) 3x3 utilize input (A, B, C) and carry output (P=A, , $Q = A \oplus B$, $R = AB \oplus C$)

### 3.4 Flexibility

Flexibility refers to the universality of reversible logic gate in realizing more logical function.

## 4. PAST WORK

In 2010, Himanshu thapliyal et.al [17] design a reversible 8-bit and 64-bit tree-based comparator using TR Gate that has the latency of O(log2(n)).But this approach is not suitable for low value of reversible parameter and not extended for n-bit reversible comparator. Furthermore another comparator design by Rangaraju et.al., [7, 8] in 2011 has shown that design has input circuit as first stage and 1-Bit comparator cell as second stage and so on. This idea is extend for n-bit reversible comparator design but it is not sufficient to optimizing the reversible parameter. In 2011, Nagamani et.al. [21] Design a reversible 1-bit reversible comparator with low value of reversible parameter but this idea is not extended for n-bit comparator.

In 2013, Ri-gui Zhou et.al. [5] design a novel 4-bit reversible comparator. It is mentioned in paper further possibilities of reducing the number of reversible gates, constant input and garbage output in the area of reversible comparator design. They proposed new gate and using this gate form various gate AND, OR, XNOR, NOT, FA, FS but not form gate NAND, NOR. For increasing order of comparator, gate increase in higher order and complexity of design increases and also garbage outputs means performance of comparator degrade.

In 2013, Hafiz Md. Hasan babu et.al. [4] design compact n-bit reversible comparator. Its design is efficient because it reduce complexity of design. Approach is sufficient and proposed various theorem and lemma for calculating n-bit reversible parameter number of gate, total delay, power. They proposed two new gate BJS and HLN. These gate form OR, AND, XOR and NOT Operation but not form FA, FS, NAND and NOR Operation. For showing expertise of reversible gate we work on this and proposed Universal gate called Inventive Gate. It's performing all logical operation NAND, NOR, AND, OR, HA, HS, FA and FS and also form efficient group-based n-bit reversible comparator and n-to-$2^n$ bit decoder with low value of reversible parameters.

## 5. UTILITY AND DESIGN ISSUES OF FRESHLY PROJECTED 4X4 INVENTIVE GATE

In this section, a new 4 x 4 reversible gate named as Inventive gate, is proposed. The block representation of inventive gate is depicted in Figure 1. The corresponding truth table drawn in Table 1. The inventive gate generates 4 outputs that are defined as follows $P = a \oplus b \oplus c$, $Q = (a \oplus b \oplus d)c \oplus b(a \oplus d)$, $R = (\overline{bc} + d)\overline{a} + bc$ and $S = \overline{bd}(a+c) + d(b + \overline{ac})$ of inventive gate.

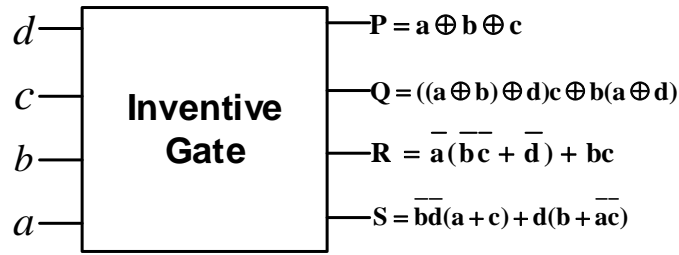

**Figure 1.** Block representation of Reversible 4 × 4 Inventive

Gate   Table 1 Reversibility of the novel Inventive gate

| INPUT | | | | OUTPUT | | | |
|---|---|---|---|---|---|---|---|
| d | c | b | a | P | Q | R | S |
| 0 | 0 | 0 | 0 | 0 | 0 | 1 | 0 |
| 0 | 0 | 0 | 1 | 1 | 0 | 0 | 1 |
| 0 | 0 | 1 | 0 | 1 | 0 | 1 | 0 |
| 0 | 0 | 1 | 1 | 0 | 1 | 0 | 0 |
| 0 | 1 | 0 | 0 | 1 | 0 | 1 | 1 |
| 0 | 1 | 0 | 1 | 0 | 1 | 0 | 1 |
| 0 | 1 | 1 | 0 | 0 | 1 | 1 | 0 |
| 0 | 1 | 1 | 1 | 1 | 1 | 1 | 0 |
| 1 | 0 | 0 | 0 | 0 | 0 | 1 | 1 |
| 1 | 0 | 0 | 1 | 1 | 0 | 0 | 0 |
| 1 | 0 | 1 | 0 | 1 | 1 | 0 | 1 |
| 1 | 0 | 1 | 1 | 0 | 0 | 0 | 1 |
| 1 | 1 | 0 | 0 | 1 | 1 | 0 | 0 |
| 1 | 1 | 0 | 1 | 0 | 0 | 0 | 0 |
| 1 | 1 | 1 | 0 | 0 | 1 | 1 | 1 |
| 1 | 1 | 1 | 1 | 1 | 1 | 1 | 1 |

### 5.1 Utility-1 of Inventive gate

The proposed Inventive gate can implement all Boolean expression and it is a universal gate. In this paper, Inventive gate is used to design the n-to-$2^n$ decoder (Shown in section 5.2) and 1-bit comparator (Shown in section 6.1). The classical operations realized using Inventive gate is shown in Fig a, b, c, d, e, f, g and h.

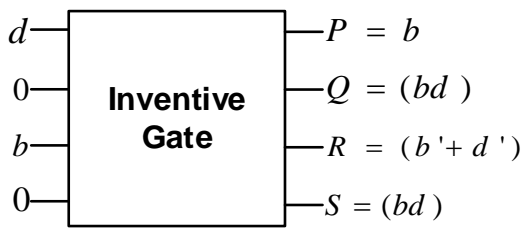

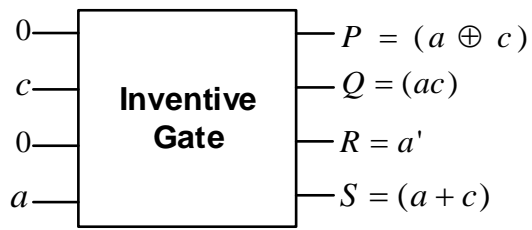

**Fig. 1 a.** AND, NAND  gates Implementation

**Fig. 1 b.** XOR, AND, NOT, OR gates Implementation

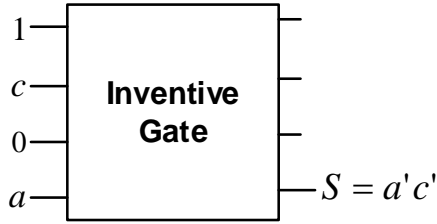

**Fig.1 c.** NOR gate Implementation

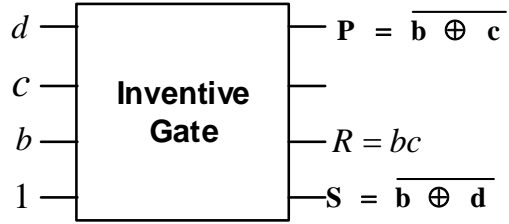

**Fig. 1 d.** XNOR, AND gates Implementation

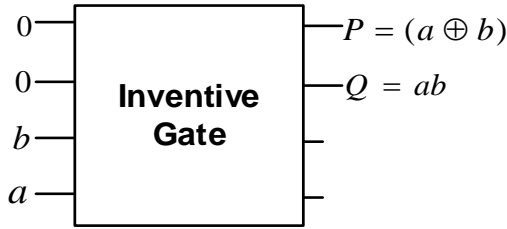

**Fig. 1 e.** Half adder Implementation

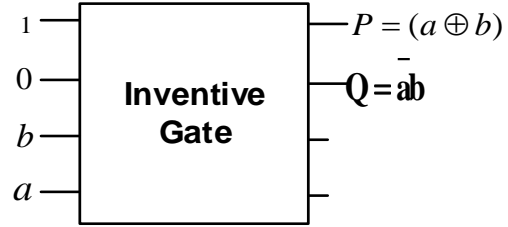

**Fig. 1 f.** Half subtraction Implementation.

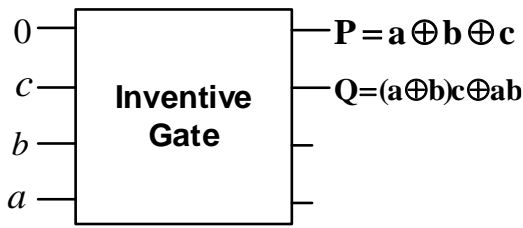

**Fig. 1 g.** Full adder Implementation

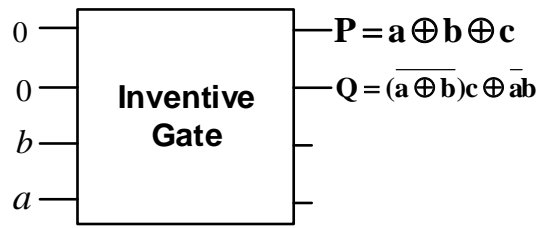

**Fig. 1 h.** Full subtraction Implementation

### 5.2. Utility-2 of Reversible inventive gate as reversible n-to-$2^n$ decoder

In this section we show the utility-2 of Inventive gate. It performs the operation of Reversible 2-to-$2^2$ decoder it consists of two approaches named as approach - 1 and approach - 2. Approach- 1 of 2-to-$2^2$ decoder generate 2 garbage output and number of gate count is 5. Whereas the Approach - 2 generate 2 garbage output but reduces number of gate count is 4 as shown in Figure 2 (b). And also implement of 3-to-$2^3$ decoder (Approach 2) generate 3 garbage output and number of gate count 7 as shown in Figure 4. Decoder [3, 11, 14] are widely used in applications like information multiplexing, 7 segment show and memory addressing.

### 5.2.1 Reversible 2-to-$2^2$ decoder and 3-to-$2^3$ decoder

Planning of Reversible 2-to-$2^2$ decoder is shown in Figure 2(b). It consists of (1 inventive gate+ 2NOT+ 1 FG+ 1TG) types of gates. It has two input marked a, b apply to inventive gate and Feynman gate (FG) respectively obtained output is passed to TG and gives output is (ab) other output of decoder is (a' b), (a b') and (a' b'). Completed cell of 2-to-$2^2$ decoder are named as I_F_T decoder cell. These cell are utilize for designing 3-to-$2^3$ decoder as shown in Figure 4.

**Lemma 5.2.1.1** An n-to-$2^n$ reversible decoder (Approach 1) may be accomplished by a minimum of $2^n + 2$ reversible gates, where n is that the range of bits and n ≥ 2

**Proof.** We ensure the above statement by mathematical induction.

For 2-to-$2^2$ decoder n = 2 decoder (Approach- 1) is made exploitation (1 inventive gate+ 2NOT+ 2 FG+ 1TG) types of gates needs a minimum of 6 ($2^n + 2$) reversible gate. Therefore the statement assert for the base value of n=2.

Accept the statement for n = y thus a y-to-$2^y$ decoder may be realized by a minimum of $2^y + 2$ reversible gates.

For (y+1)-to-$2^{(y+1)}$ decoder is made exploitation y- to -$2^y$ decoder and $2^y$ FRG Gate. So the Total number of gates needed to construct a (y+1) - to- $2^{(y+1)}$ decoder is a minimum of $2^y + 2 + 2^y = 2^{y+1} + 2$

So the statement hold for n = y+1

**Lemma 5.2.1.2** An n-to-$2^n$ reversible decoder (Approach - 2) may be accomplished by a minimum of $2^n+1$ reversible gates, where n is that the range of bits and n$\geq$ 2

**Proof.** We ensure the above statement by mathematical induction.

For 2-to-$2^2$ decoder n = 2 decoder (Approach -2) is made exploitation (1 inventive gate+ 2NOT+1 FG+1TG) type of gates needs a minimum of 5 ($2^n+1$) reversible gate. Therefore the statement assert for the base value of n=2.

Accept the statement for n = y thus a y-to-$2^y$ decoder may be realized by a minimum of $2^y+1$ reversible gate.

For (y+1)-to-$2^{(y+1)}$ decoder is made exploitation y-to-$2^y$ decoder and $2^y$ FRG Gate. So the total range of gates needed to construct a (y+1)-to-$2^{(y+1)}$ decoder is a minimum of.

$2^y+1+2^y = 2^{y+1}+1$

So the statement hold for n = y+1 and more powerful than approach 1 because it less number of gate count.

**Lemma 5.2.1.3** An n-to-$2^n$ reversible decoder (Approach -2) may be accomplished by a minimum of n garbage output, where n is that the range of bits and n$\geq$ 2

**Proof**. We ensure the above statement by mathematical induction.

For 2-to-$2^2$ decoder has garbage output 2 (n) for the base value of n = 2. For higher order decoder 3-to-$2^3$ garbage output 3 (n) for n=3

Accept the statement for n = k thus a k: $2^k$ decoder may be realized by a minimum of k garbage output.

International Journal of VLSI design & Communication Systems (VLSICS) Vol.5, No.5, October 2014

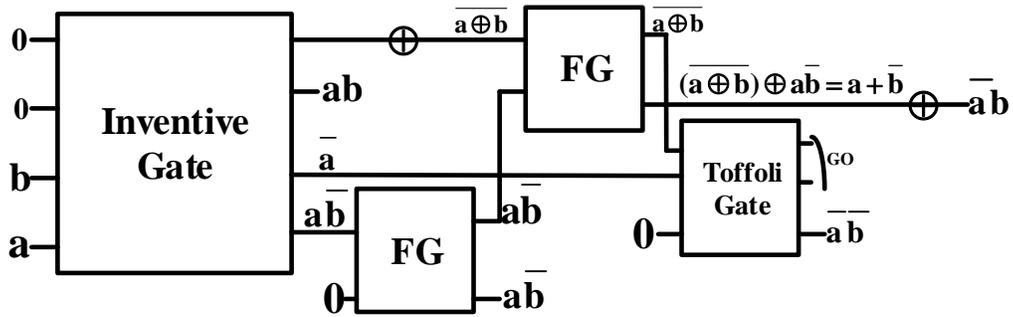

**Fig. 3 a.** Approach- 1

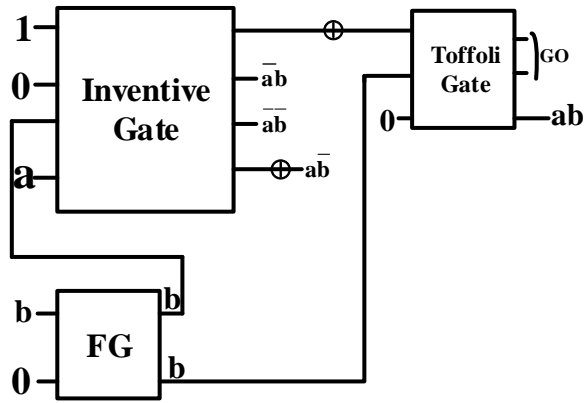

**Fig. 3 b**. Approach – 2

**Figure 2.** Two approaches of the Novel 2- to- $2^2$ decoder

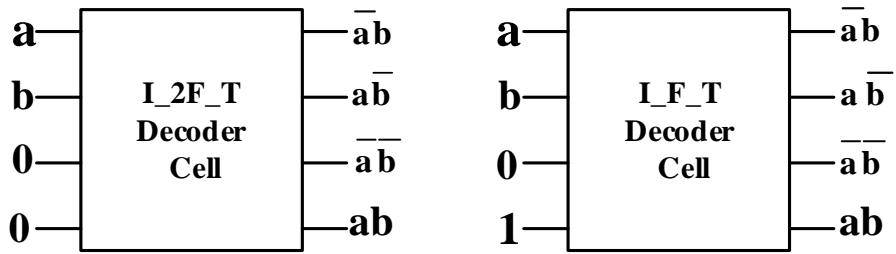

**Figure 3.** Approach- 2 of proposed reversible 2-to-$2^2$ decoder cell.

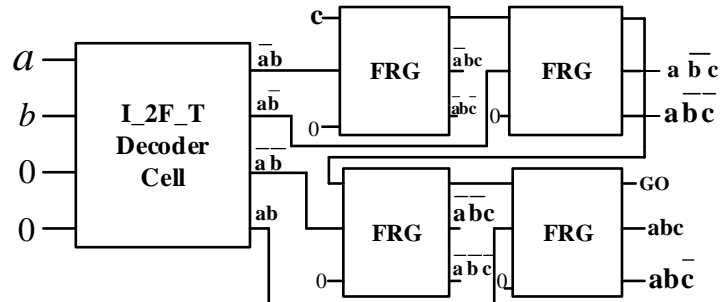

**Fig. 4 c.** Approach- 1

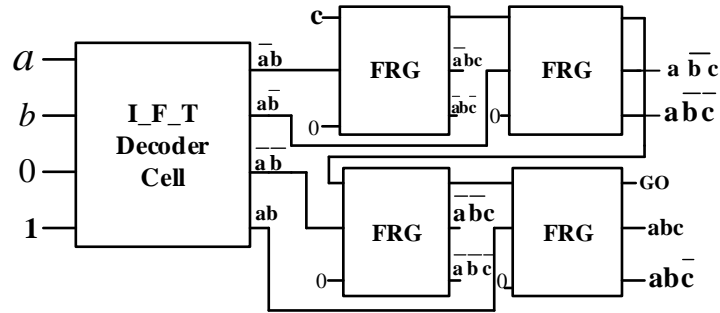

**Fig. 4 d.** Approach- 2

**Figure 4.** Two Approaches of the proposed 3 -to- $2^3$ decoder

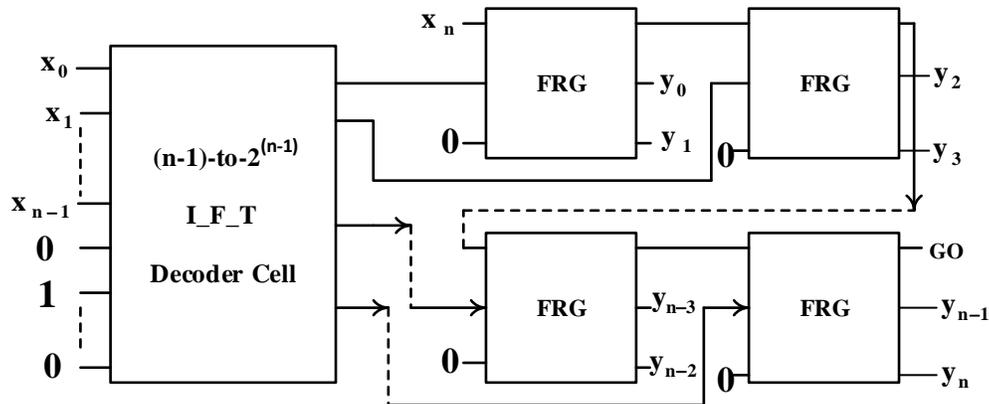

**Figure 5.** Approach- 2 of proposed reversible n-to-$2^n$ decoder.

## 6. NOVEL DESIGN OF GROUP-BASED REVERSIBLE N-BIT BINARY COMPARATOR

Following section we have to project a low value of group-based reversible n-bit comparator.

### 6.1 Novel 1-bit reversible comparator cell design

Novel 1-bit group-based reversible comparator structure uses 3 gates of 2 types (1 inventive+ 2 NOT) .It is true that $a\bar{b}$ for (a>b), $\bar{a}b$ for (a<b) and $\overline{a \oplus b}$ for (a=b). The proposed architecture has constant inputs of 2, Garbage outputs of 1 and number of gate count 3. The structure of 1-bit comparator modelled as I_N Comparator cell. As shown in Figure 6 (a)

## 6.2 Novel 1- bit Match and larger reversible comparator cell design

Novel 1-bit match and larger comparator structure use 4 Gate of 4 types (1 TR+ 1 NOT+1 BME+1 FG) .First reversible TR gate use (n- 1)$^{th}$ bits of two logic input a and b it produce two significant output $\overline{a \oplus b}$ and $a\overline{b}$ first output is not utilize is called garbage output. Two more n$^{th}$ bit input P$_n$ and Q$_n$ of previous comparator result is applied to reversible BME Gate it produce two significant output Q$_{n-1}$ = Q$_n$ $\overline{a \oplus b}$ and Q$_n$ $(a\overline{b})$ only Q$_{n-1}$ output save and other output Q$_n$ (a b') is applied to another reversible Feynman gate (FG) and another n$^{th}$ bit input P$_n$ input is selecting give one significance output P$_{n-1}$ = Q$_n$ $(a\overline{b})$ ⊕ P$_n$ these two output P$_{n-1}$ and Q$_{n-1}$ is utilizing further for designing lesser logic comparator design. The proposed architecture has constant input of 2, garbage output of 4 and number of gate count 4 (1+1+1+1). The structure of Larger and Match design modelled as TR_BME_FG Comparator cell. As shown in Figure 6 (b)

## 6.3 Novel 1-bit smaller reversible comparator cell design

Novel single bit smaller reversible comparator structure use 4 gate of 2 type (3 FG + 1 NOT).These structure utilize two input P and Q and give two significant output P and (P⊗Q) second output (P ⊕ Q) is passed to NOT gate gives $\overline{P \oplus Q}$ other output is P and Q for match operation use (P), Larger operation (Q) and lesser operation $\overline{P \oplus Q}$. The novel design has constant input of 1, Garbage output of 0 and Number of gate count 3 (1+1+1). The structure of lesser cell modelled as F_F Comparator cell. As shown in Figure 6 (c)

## 6.4 Novel Efficient 2-bit reversible comparator design

Design methodology of 2-bit group- based comparator cell used three cell pervious section named I_N Comparator cell, TR_BME_FG Comparator cell and F_F Comparator cell these cell are connect and gives significant output P,Q and R for Match, Smaller and Larger logic. The novel design is shown in Figure. 6 (d)

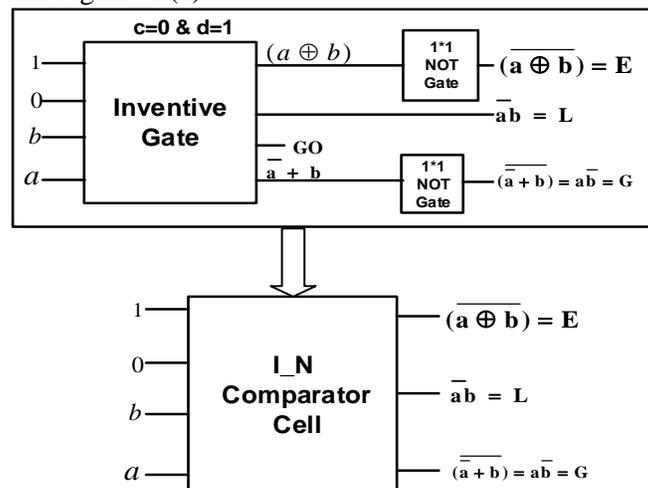

**Fig. 6 a.**

International Journal of VLSI design & Communication Systems (VLSICS) Vol.5, No.5, October 2014

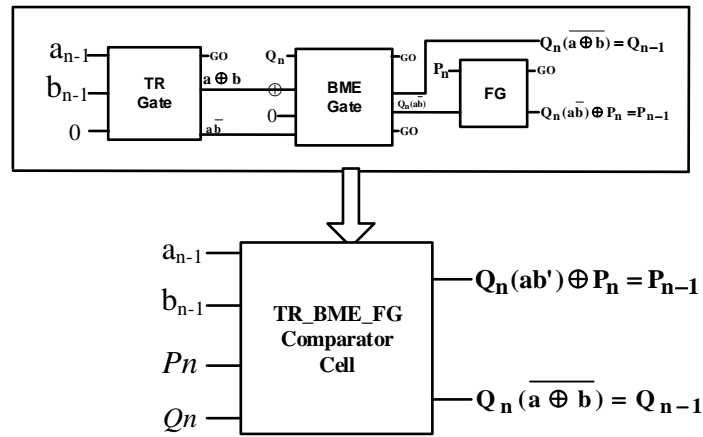

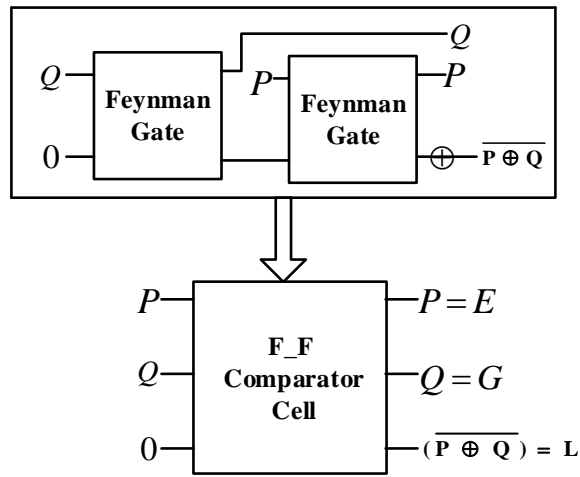

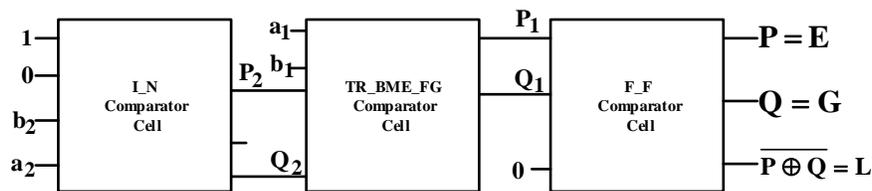

**Fig. 6d.**
**Figure 6**. Reversible group-based 2-bit comparator design

**a.** Reversible 1-bit Comparator.
**b.** Match and Larger Cell design using TR_BME_FG Cell.
**c.** Smaller Comparator Cell design using F_F Cell.
**d.** Reversible 2-bit Comparator.

## 6.5 Novel architecture of Reversible 8-bit group-based Comparator

Following the extremely comparable analogous approach kept in mind then we are design structure of 8-bit group-based Comparator module. The main projected structure for the 8-bit

Comparator is anticipated in Figure 7. The proposed architecture has constant input of 17(1×2+2×7+1), Garbage output of 29 (1+ 4×7+ 1×0) and Number of gate count 34 (1×3+4×7+1×3)

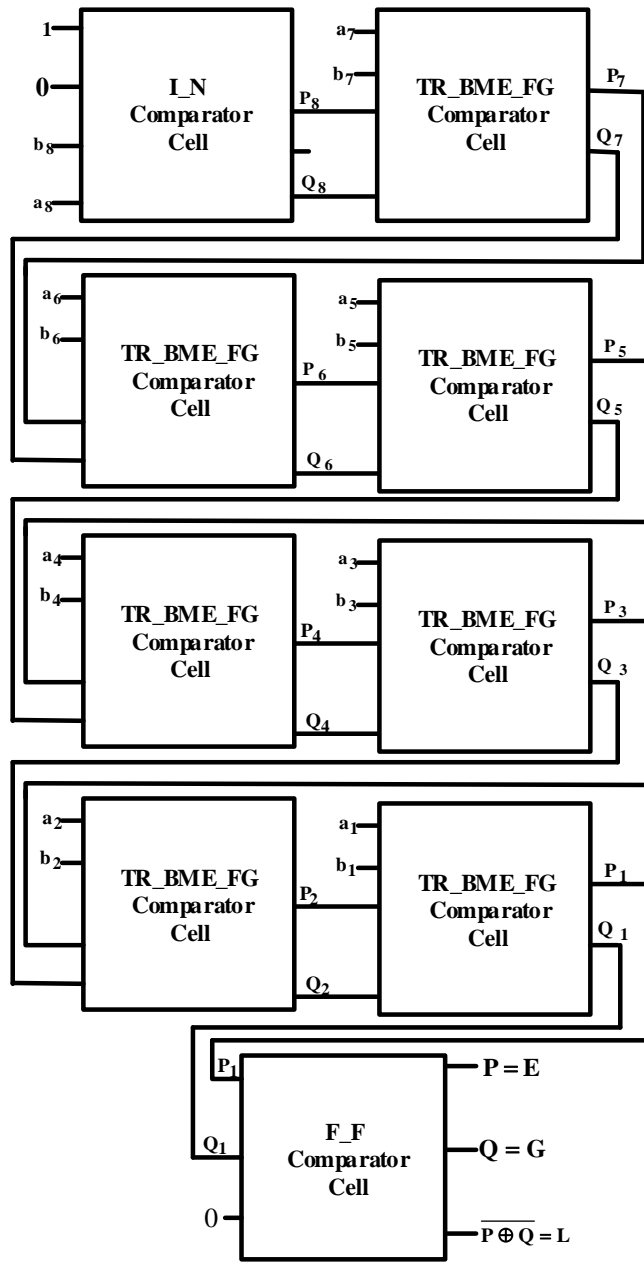

Figure 7. Novel architecture of Reversible 8-bit Comparator

## 6.6 Novel architecture of Reversible 32-bit Group-based Comparator structure.

For the implementation methodology of 32-bit group-based comparator design use (1 I_N Comparator Cell+ 31 TR_BME_FG Comparator cell+1 F_F Comparator cell) .The main projected structure for the 32-bit binary group-based comparator cell is anticipated in Figure 8.

The proposed architecture has constant input of 65 (1×2+ 2×31+ 1×1), Garbage output of 125 (1×1+ 4×31+ 1×0) and Number of gate count 33 (1+ 31+ 1)

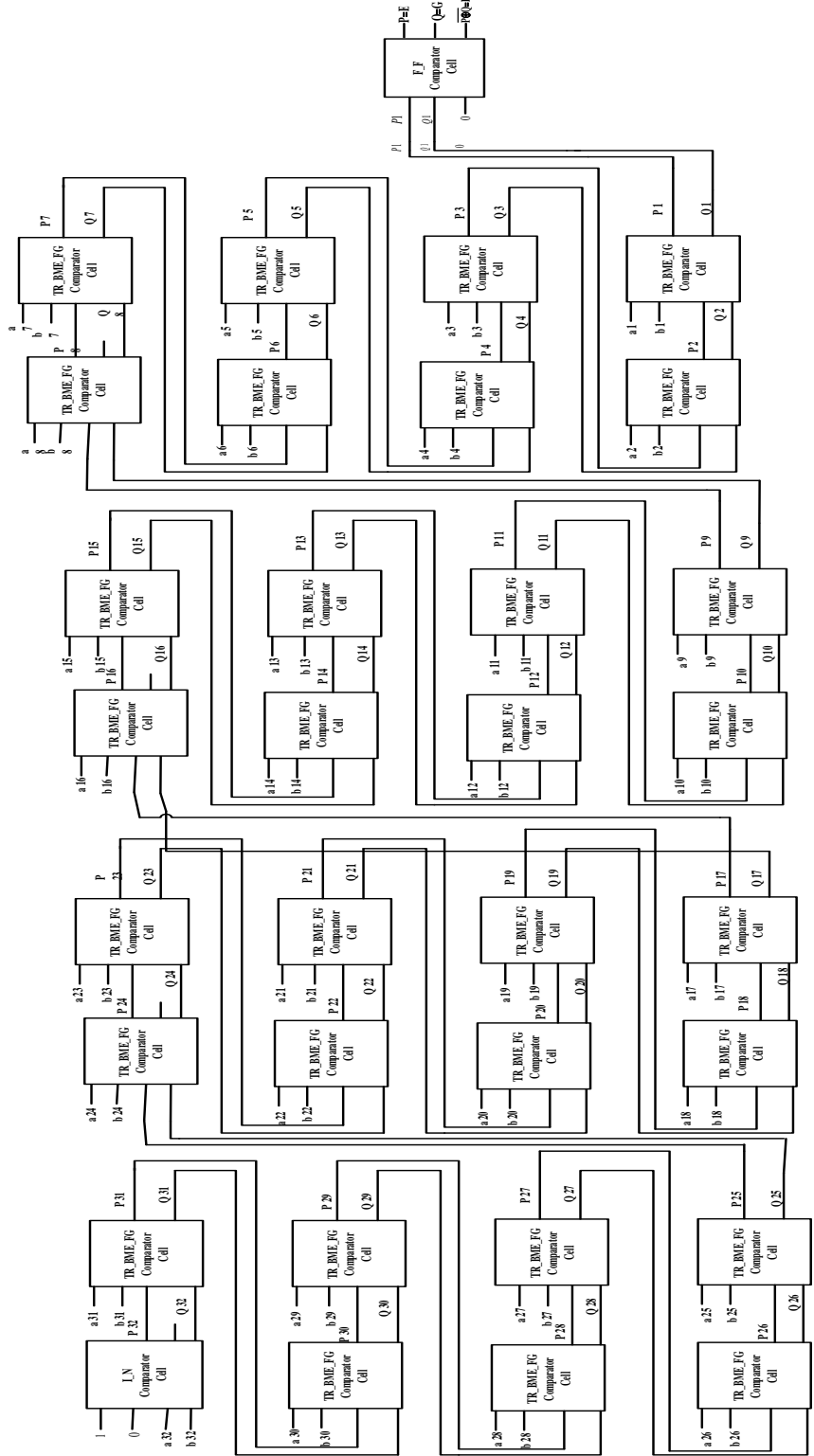

**Figure8.** Novel architecture of group-based reversible 32-bit comparator.

## 6.7 Novel architecture of Reversible n- bit group- based Comparator structure

Following the extremely comparable analogous approach we are proposing n- bit group-based comparator structure. First idea is replenishment to the main projected structure for the n-bit group-based comparator structure is anticipated in Figure 9. For designing 32-bit tree comparator cell consist of $n^{th}$ single bit to I_N comparator cell and $(n-1), (n-2), (n-3)………..(n-31)^{th}$ single bit apply to TR_BME_FG Comparator cell where n is 32 for 32-bit comparator. This concept is apply for n-bit group-based comparator it consist of $n^{th}$ single bit to I_N comparator cell and $(n-1), (n-2), (n-3)………..(n-31)………….(n-y)^{th}$ single bit apply to TR_BME_FG comparator cell where $y=(n-1)$

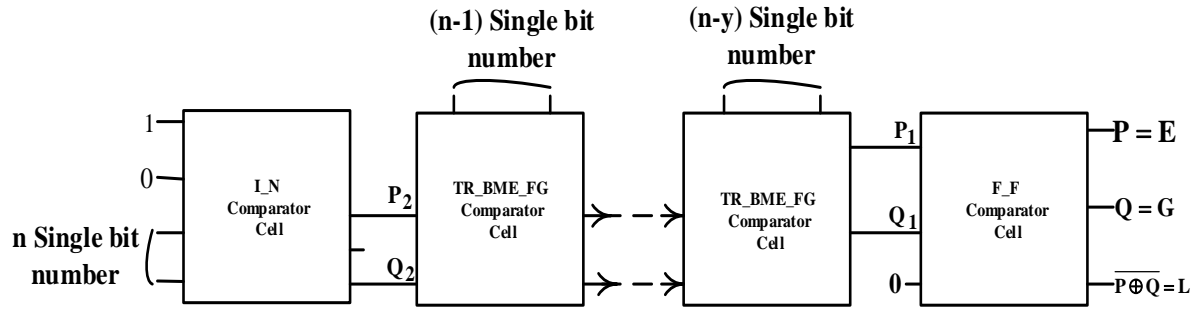

**Figure 9.** Novel n- bit group- based reversible Comparator

The steps to style a n-bit group-based comparator are describe in Algorithm 1.

**Algorithm1.** Group- based n-bit Reversible Comparator

**(1)** Pick up I_N Comparator cell and acquire input $a_n$ and $b_n$

**(2)** Pick up TR_BME_FG Comparator cell and takes $(n-1)^{th}$ bits of 2-binary variety $a_{n-1}$ and $b_{n-1}$ and 2 additional input Pn and Qn from previous comparison result.

**(3)** TR_BME_FG Comparator cell 2-output Pn-1 and Qn-1 that indicates whether the given binary number are equal to one another or larger.

**(4)** Pick up F_F Comparator cell and takes input P, Q it generates output such as P (For Equal) L(Less) and G (For Greater).

**Lemma 6.7.1** An n-bit group-based reversible comparator may be accomplished by a 6+4 (n-1) number of gate, 1+4 (n-1) garbage output, where n is the range of bits and n 2

**Proof:** We ensure the above statement by mathematical induction.

An n-bit reversible comparator has number of gate 6+ 4 (n-1). For realization of 2-bit and 8-bit reversible comparator setting n=2, 8 alternatively gives 10 and 34 number of gate.

An n-bit reversible comparator has garbage output 1+ 4 (n-1) .For realization of 2-bit and 8-bit reversible comparator setting n=2, 8 alternatively give 5 and 29 number of garbage output.

**Lemma 6.7.2** An n-bit group-based reversible comparator requires (85.816n – 78.989) µW power, where n is the range of bits and n ≥ 2

**Proof:** Novel n-bit group-based comparator uses I_N Comparator cell, (n– 1) TR_BME_FG Comparator cell and one F_F Comparator cell. Total power relished by following specified equation

Total Power (P) = $P_{I\_N\ Cell}$ + (n-1) $P_{TRE\_BME\_FG\ Cell}$ + $P_{F\_F\ Cell}$

All Comparator cell implement in MOS transistor and finding power using T-Spice tool for 90nm technology node. We have computed the power of I_N comparator cell, TRE_BME_FG Comparator cell and F_F Comparator cell that are 3.358, 85.816 and 3.470 µW respectively. Now the total power (P) of an n-bit group-based comparator as below

P= {3.35 + (n – 1) 85.81+ 3.4702} µW

= (3.35 +85.81n – 85.81 + 3.4702) µW

= (85.81n – 78.98) µW

**Lemma 6.7.3** An n-bit group-based reversible comparator may be accomplished by a Timing delay (T) of (115.010 n – 100.854) ns, where n is the range of bits and n ≥ 2

**Proof:** Novel n-bit group-based Comparator uses I_N Comparator cell, (n– 1) TR_BME_FG Comparator cell and one F_F Comparator cell .Total delay (T) relished by following specified equation

Total delay (T) = $T_{I\_N\ Cell}$ + (n – 1) $T_{TRE\_BME\_FG\ Cell}$ + $T_{F\_F\ Cell}$

All Comparator cell implement in MOS transistor and finding delay using T-Spice tool for 90nm technology node. We have computed the power of I_N comparator cell, TRE_BME_FG Comparator cell and F_F Comparator cell that are 12.51, 115.010 and 1.608 ns respectively. Now the delay of an n-bit group-based comparator as below

T= {12.51+ (n –1) 115.010 + 1.608} ns
 = (12.51+ 115.010 n – 115.010 + 1.608) ns
 = (115.010 n – 100.854) ns

**Algorithm- 2 Reversible group-based n-bit comparator**

*Design a n-bit reversible binary comparator when ($n \geq 2$) with minimum number of gates and garbage output.

**Begin**
**Step 1.** Pick up I_N comparator cell and pick input $X_n$ and output $Y_n$ for two n-bit number
**1.** $X_n[1] = a_n$   //$n^{th}$ input of $a_n$
**2.** $X_n[2] = b_n$   //$n^{th}$ input of $b_n$
**3.** $X_n[3] = 0$
**4.** $X_n[4] = 1$
**5.** If        $X_n[1] < X_n[2]$ then
                 $Y_n[2] = R_{yn} = 1$
**6. Else** if    $X_n[1] > X_n[2]$ then
                 $Y_n[4] = Q_{yn} = 1$

**7.** Else      $Y_n[1] = P_{yn} = 1$
End if
**Step 2.** For TR_BME_FG and F_F Comparator cell, Level of input and output are considered to be $Y_n$ and $Y_{n-1}$ respectively
Loop
**8.** For j = n - 1 to 1
Pick up TR_BME_FG Comparator cell $W_J$
If j = n-1 then
$W_J[1] = X_n[4] = Q_{Yn}$
$W_J[2] = X_n[1] = P_{Yn}$
$W_J[3] = b_{n-1}$
$W_J[4] = a_{n-1}$
Else
$W_J[1] = W_{J-1}[2] = Q_{Yn-1}$
$W_J[2] = W_{J-1}[1] = P_{Yn-1}$
$W_J[3] = a_j$              //$(n-1)^{th}$ input of $a_n$
$W_J[4] = b_j$              //$(n-1)^{th}$ input of $b_n$
End if
End loop
**9. Step 3.** Pick up F_F Comparator cell and pick input V and output Z
$V[1] = W[1]$
$V[2] = W[2]$
$V[3] = 0$
$Z[1] = W[1]$
$Z[2] = [W[1] \oplus W[2]] \oplus 1$
$Z[3] = W[2]$
**10.** End

# 7. MOS TRANSISTOR IMPLEMENTATION OF DIFFERENT COMPARATOR CELL

In this section implement different comparator cell in MOS transistor with minimum MOS transistor count. MOS transistor implementations of 2-bit comparator uses 14 MOS transistors,

International Journal of VLSI design & Communication Systems (VLSICS) Vol.5, No.5, October 2014

TR_BME_FG Comparator cell required 18 MOS transistor and for F_F comparator cell require 4 MOS transistor. These cells are connected for required 2-bit, 8-bit and 64-bit reversible comparator operation and analyzed in terms of power consumption, delay and power delay product (PDP).

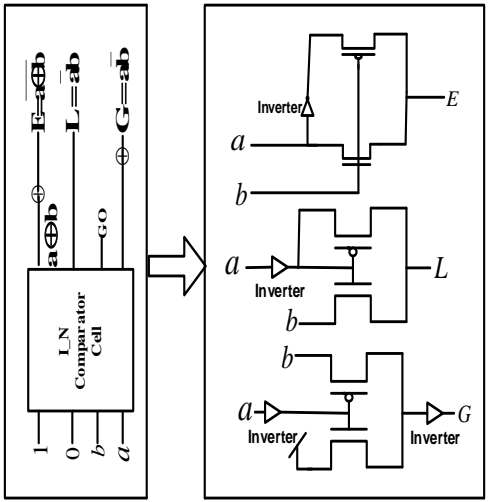
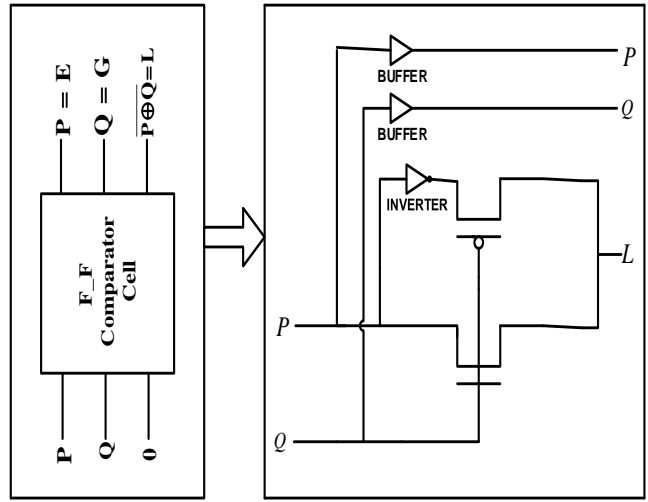

**Figure 10.** MOS transistor implementation of 1-bit Comparator Cell

**Figure 11.** MOS transistor implementation of F_F Comparator Cell

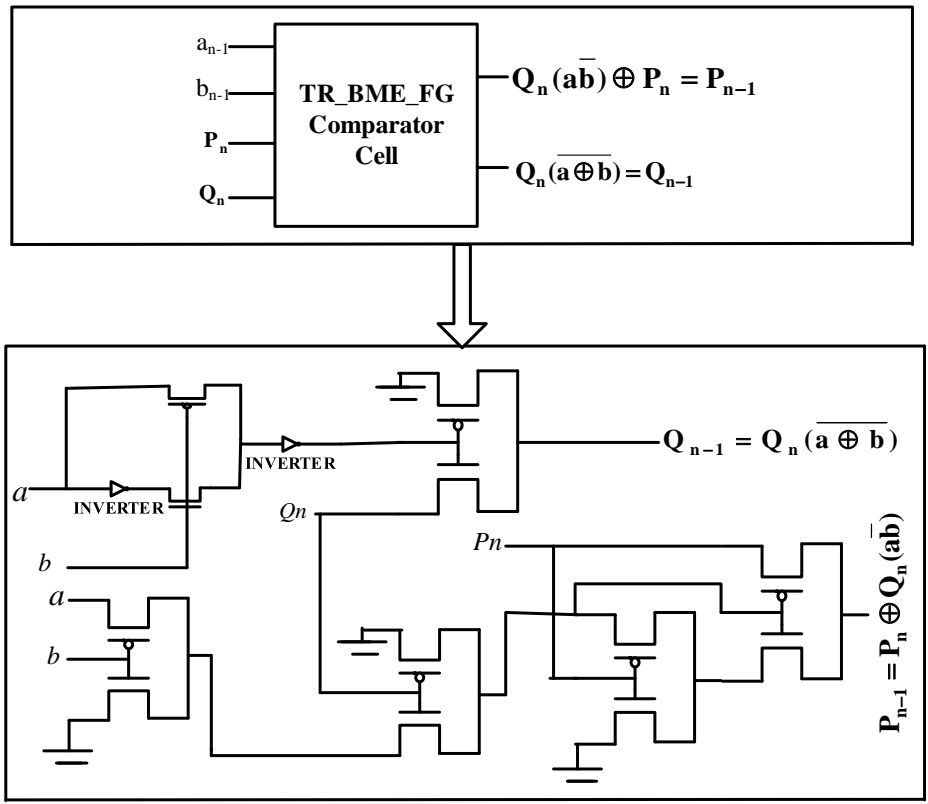

**Figure 12.** MOS transistor implementation of TR_BME_FG Comparator Cell

## 8. SIMULATION RESULT AND DISCUSSION

Novel design of 2-bit, 8-bit, 64-bit and n-bit group-based reversible comparator is implemented in T-Spice and optimized the speed, power for appropriate W/L ratio using 90nm technology node. Individual cell performance parameters analyze for reducing input voltage and finding power consumption, delay etc.

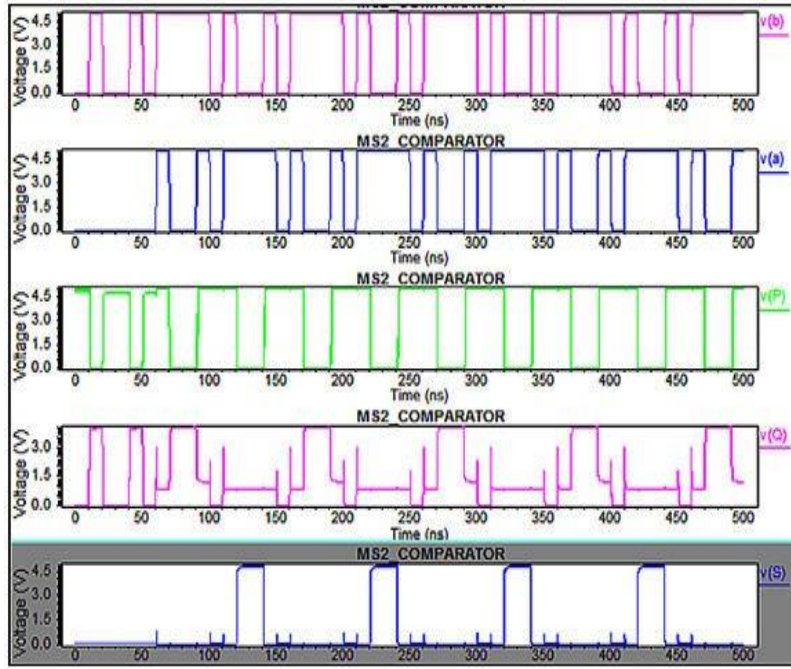

**Figure 13.** Simulation output of 1-bit comparator.

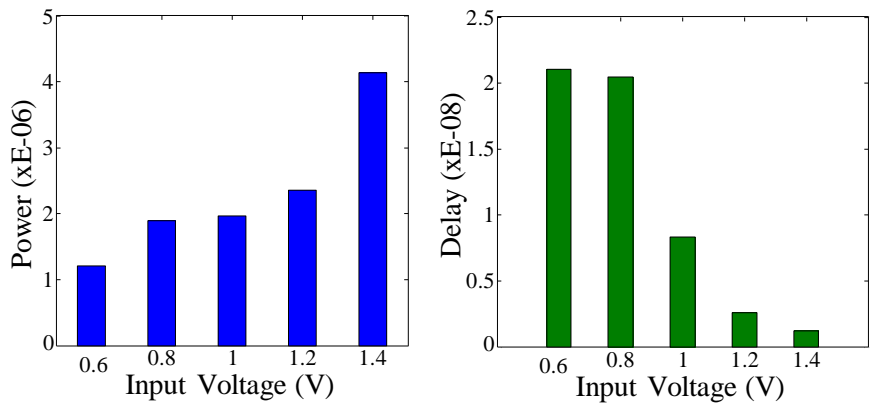

**Figure 14.** Power and delay Comparison at different input voltage of 1-bit comparator.

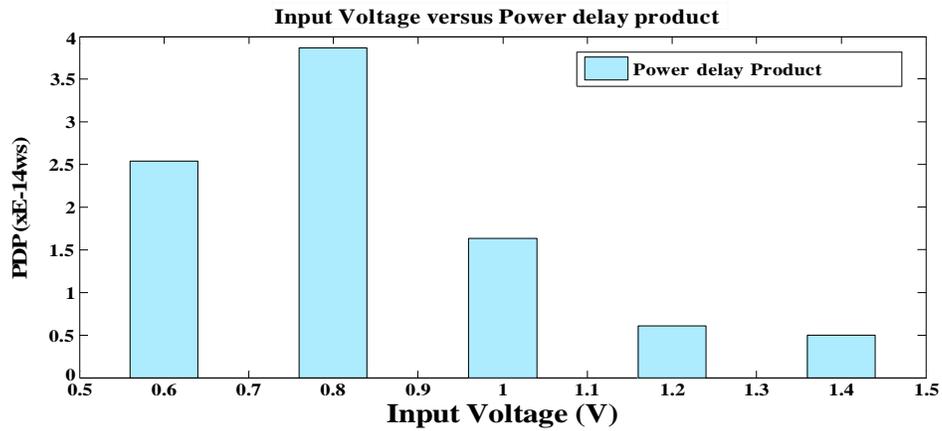

**Figure 15.** Input voltage versus Power delay product (PDP) of 1-bit Comparator.

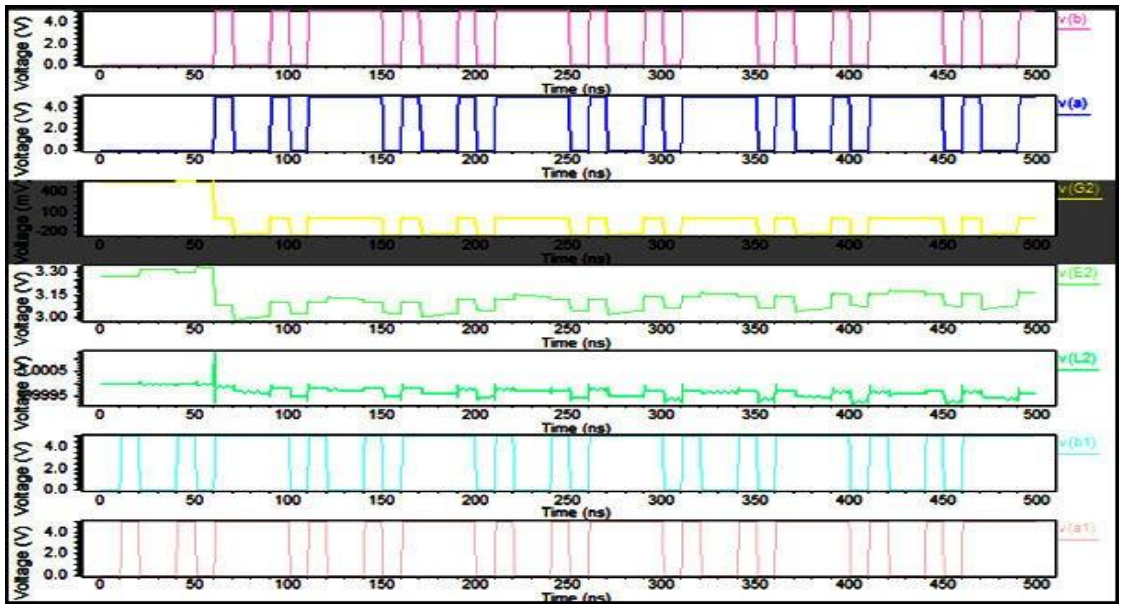

**Figure 16.** Simulation output of 2-bit comparator.

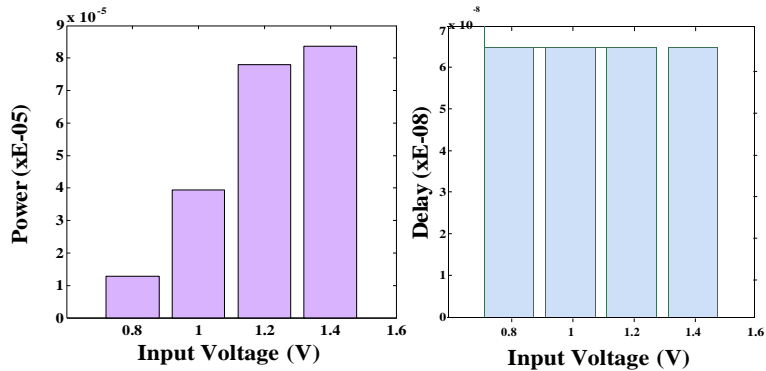

**Figure 17.** Power and delay Comparison at different Input voltage of 2-bit comparator.

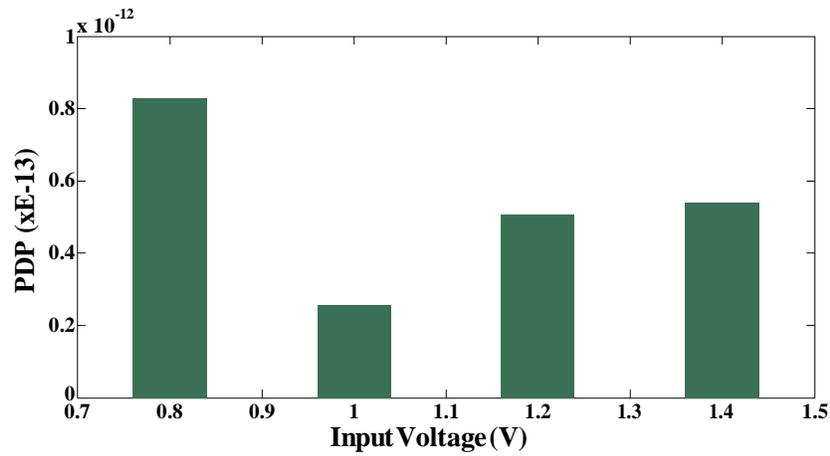

**Figure 18.** Input voltage versus Power delay product (PDP) of 2-bit Comparator.

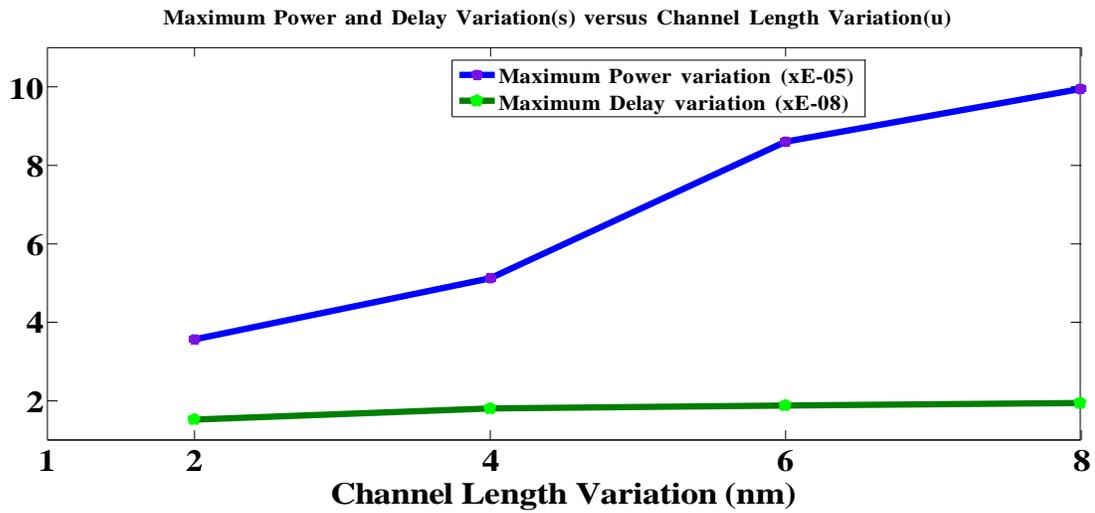

**Figure 19.** Channel Length Variation versus power and delay of 2-bit Comparator

Table 2. Comparison between anticipated and existing in terms of expertise of reversible gate

| RGate | Gate can be used as n-bit comparator | Single Gate can be used as Full subtractor | Gate can be used as n-to-$2^n$ decoder |
|---|---|---|---|
| Anticipated Inventive Gate | Y | Y | Y |
| Available Circuit [4] | Y | N | N |
| Available Circuit [3] | N | N | Y |
| Available Circuit [12] | N | N | N |
| Available Circuit [13] | N | N | N |
| Available Circuit [15] | N | N | N |
| Available Circuit [6] | Y | N | N |
| Available Circuit [5] | Y | Y | N |

Table 3. Comparison between anticipated and existing style comparator in terms of NOG, GO, CI, power and delay

| Methods | NOG | GO | CI | Power (µW) | Delay (ns) |
|---|---|---|---|---|---|
| Anticipated work | 6+4(n-1) | 1+4(n-1) | 1+2n | (85.81n − 78.98) | (115.010n − 100.854) |
| Thaplial et al. [17] | 9n | (6n −6) | -- | (268.23n −239.2) | {0.23 x log2(n) + 0.1} |
| Vudadha et al. [18] | (4n −2) | (5n −4) | -- | (122.36n −60.36) | {0.09*log2(n) + 0.2} |
| Rangaraju et al. [7] | (7n −4) | (5n −4) | -- | (182.53n +76.55) | (0.2n − 0.16) |
| Hafiz Md. Hasan Babu [4] | 3n | (4n-3) | 3 | (117.76n −32.94) | (0.15n − 0.03) |

NOG-number of gate, GO-Garbage output, CI-Constant input, Y=Yes, N=No

Table 4 Comparison between anticipated and existing style Comparator in terms of garbage output and constant input

| Methods | 8-bit Comparator | | 16-bit Comparator | | 32-bit Comparator | |
|---|---|---|---|---|---|---|
| | GO | CI | GO | CI | GO | CI |
| Anticipated work | 29 | 17 | 61 | 33 | 125 | 65 |
| Rangaraju et al.[7] | 36 | 23 | 76 | 47 | 156 | 95 |
| Thapliyal et al. [17] | 42 | -- | 90 | -- | 186 | -- |
| Vudadha et al.[18] | 36 | -- | 76 | -- | 156 | -- |
| Hafiz Md. Hasan Babu [4] | 29 | -- | 61 | -- | 125 | -- |
| Morrison et al.[20] | 39 | -- | 79 | -- | 159 | -- |
| % improvement w.r.t [7] | 19.44 | 26.08 | 19.73 | 29.78 | 19.87 | 31.57 |
| % improvement w.r.t [17] | 30.95 | -- | 32.22 | -- | 32.79 | -- |
| % improvement w.r.t [18] | 19.44 | -- | 19.73 | -- | 19.87 | -- |
| % improvement w.r.t [20] | 25.64 | -- | 22.78 | -- | 21.13 | -- |

## 9. CONCLUSION AND FUTURE WORK

This paper is mainly focused on novel design of the reversible 4x4 inventive Gate. It utilized as 1-bit, 2-bit, 8-bit, 32-bit and n-bit group-based binary comparator and n-to-$2^n$ decoder. Moreover, the low value styles of reversible parameter have been established for the proposed circuits. For example, and n-to-$2^n$ decoder uses at least $2^n + 1$ reversible gates and its n garbage output: an n-bit comparator utilize 6+4 (n - 1) number of reversible gate, 1+4 (n-1) garbage output. The proposed group- based comparator achieves the improvement of 19.87% in terms of garbage output and 31.57% in terms of constant input over the existing one [7]. Simulation of the novel comparator circuits have shown that it works correctly and finding parameter power and delay. The proposed circuits will be useful for implementing the ALU and control unit of processor.

**ACKNOWLEDGMENT**

Authors would like to acknowledge Research Cum Teaching Fellowship through TEQIP-II provide by World Bank for the financial support for this work.

**REFERENCES**

[1] J R.Landauer, "Irreversibility and Heat Generation in the Computational Process", IBM Journal of Research and Development, pp: 183-191, 1961.
[2] C H Bennett, "Logical Reversibility of Computation," IBM Journal of Research and Development, vol. 17, no. 6, pp. 525-532, November 1973.
[3] Lafifa Jamal, Md. Masbaul Alam, Hafiz Md. Hasan Babu, "An efficient approach to design a reversible control unit of a processor" Elsevier Sustainable Computing: Informatics and Systems pp: 286-294, 2013
[4] Hafiz Md. Hasan Babu, Nazir Saleheen, Lafifa Jamal, Sheikh Muhammad Sarwar,Tsutomu Sasao "Approach to design a compact reversible low power binary comparator" IET Computers & Digital Techniques" Vol. 8, Iss. 3, pp. 129–139 doi: 10.1049/iet-cdt.2013.0066, 2014
[5] Ri-gui Zhou, Man-qun Zhang, QianWu •Yan-cheng Li "Optimization Approaches for Designing a Novel 4-Bit Reversible Comparator" Springer International jounal of theoretical physics DOI 10.1007/s10773-012-1360-y pp:559-575, 2013
[6] Farah Sharmin, Rajib Kumar Mitra, Rashida Hasan, Anisur Rahman "Low cost reversible signed comparator" International Journal of VLSI design & Communication Systems" Vol.4, No.5,pp:19-33,2013
[7] Rangaraju H G, Vinayak Hegdeb, Raja K B, Muralidhara K N "Design of Efficient Reversible Binary Comparator" Elsevier International Conference on Communication Technology and System Design Procedia Engineering pp:897-904, 2012
[8] Rangaraju, H.G., Hegde, V., Raja, K.B., Muralidhara, K.N. "Design of low power reversible binary comparator". Proc. Engineering (ScienceDirect), 2011
[9] Bahram Dehghan, Abdolreza Roozbeh, Jafar Zare "Design of Low Power Comparator Using DG Gate" Scientific research ciruits and systems doi.org/10.4236/cs.2013.51002pp: 7-12, 2014
[10] Pallavi Mall, A.G.Rao, H.P.Shukla "Novel Design of Four-Bit Reversible Numerical Comparator" International Journal of Advanced Research in Computer and Communication Engineering Vol. 2, Issue 4 pp: 1808-1807, 2013
[11] Neeta Pandey, Nalin Dadhich, Mohd. Zubair Talha "Realization of 2-to-4 reversible decoder and its applications" International Conference on Signal Processing and Integrated Networks (SPIN) pp: 349-353,2014
[12] Md. M. H Azad Khan, "Design of Full-adder With Reversible Gates", International Conference on Computer and Information Technology, Dhaka, Bangladesh, 2002, pp. 515-519

[13] Hafiz Md. Hasan Babu, Md. Rafiqul Islam, Syed Mostahed Ali Chowdhury and Ahsan Raja Chowdhury ,"Reversible Logic Synthesis for Minimization of Full Adder Circuit", Proceedings of the EuroMicro Symposium on Digital System Design(DSD'03), 3-5 September 2003,
[14] Neeraj Kumar Misra, Subodh Wairya, Vinod Kumar Singh "Preternatural Low-Power Reversible Decoder Design in 90 nm Technology Node" International Journal of Scientific & Engineering Research, Volume 5, Issue 6, pp: 969-978, June 2014
[15] J.W. Bruce, M.A. Thornton,L. Shivakumariah, P.S. Kokate and X.Li, "Efficient Adder Circuits Based on a Conservative Logic Gate", Proceedings of the IEEE Computer Society Annual Symposium on VLSI(ISVLSI'02),April 2002, Pittsburgh, PA, USA, pp 83-88.
[16] Lafifa Jamal, Md. Masbaul Alam Polash "On the Compact Designs of Low Power Reversible Decoders and Sequential Circuits" Springer LNCS7373, pp. 281–288, 2012.
[17] Himanshu Thapliyal, Nagarajan Ranganathan and Ryan Ferreira "Design of a Comparator Tree Based on Reversible Logic" 10th IEEE International Conference on Nanotechnology Joint Symposium with Nano Korea pp:1113-1116, 2010
[18] Vudadha, C.Phaneendra, P.S.,Sreehari,V.Ahmed,S.E.Muthukrishnan, N.M.Srinivas "Design of prefix-based Optimal reversible comparator" IEEE Computer Society Annual Symp. On VLSI pp. 201–206
[19] Nagamani, A.N., Jayashree, H.V., Bhagyalakshmi, H.R.: 'Novel low power comparator design using reversible logic gates', Indian J. Comput. Sci. Eng.vol 2, (4), pp. 566–574, 2011
[20] Morrison, M., Lewandowski, M., Ranganathan "Design of a tree-based comparator and memory unit based on a novel reversible logic structure" IEEE Computer Society Annual Symp. on VLSI, pp. 331–336, 2012
[21] Neeraj Kumar Misra, Subodh Wairya, Vinod Kumar Singh "An Inventive Design of 4*4 Bit Reversible NS Gate" IEEE International Conference on Recent Advances and Innovation in Engineering (ICRAIE-2014), pp: 1-6, 2014 doi no. 10.1109/ICRAIE.2014.6909323